\documentstyle[pra,aps,epsf]{revtex}

\begin{document}
\title{Antihydrogen-Hydrogen annihilation at sub-kelvin temperatures}
\author{A. Yu. Voronin}
\address{P. N. Lebedev Physical Institute\\
53 Leninsky pr.,117924 Moscow, Russia}
\author{J. Carbonell}
\address{Institut des Sciences Nucl\'eaires\\
53, Av. des Martyrs, 38026 Grenoble, France}
\maketitle

\begin{abstract}
Annihilation cross-section of ultra low-energy atomic antihydrogen ($%
E\le10^{-5}$ a.u.) on atomic hydrogen are calculated within a quantum
mechanical coupled channels approach. The results differ from the
extrapolations of semiclassical models to low energies. The main features of
the observables are found to be determined by a family of $H\bar{H}$
nearthreshold metastable states.
\end{abstract}


\section{INTRODUCTION}

The projects of synthesis and spectroscopy of ultracold atomic antihydrogen
in traps \cite{AD} require theoretical calculations of the low energy
matter-antimatter inelastic collisions rates. We present here a quantum
mechanical model of atomic antihydrogen ($\overline{H}$) and atomic hydrogen
($H$) interaction, describing inelastic collisions $(\overline{H}%
+H\rightarrow Pn^{\ast }+e^{-}e^{+})$ at energies less than $10^{-5}$a.u.

We developed a unitary coupled channel model which incorporates the main
physical inputs. Our results for the annihilation cross-section will be
compared to those provided by the semiclassical methods \cite%
{Morgan,Morgan1,SWS}. We will show that in our approach the low energy
properties are determined by a rich spectrum of nearthreshold S-matrix
singularities generated by long range van der Waals potential.


\section{THE FORMALISM}

We develope the effective potential approach to the low energy rearrangement
collisions, used by the authors for $H\bar{p}$ scattering \cite{PRA}. The
details of the formalism would be published elsewhere, here we give only
very brief description.

The wave-function is represented as a sum of two components: 
\begin{eqnarray*}
\Phi &=&\Phi _{1}+\Phi _{2} \\
\Phi _{1} &=&(1-\widehat{F})\Phi \\
\Phi _{2} &=&\widehat{F}\Phi
\end{eqnarray*}
Here $\widehat{F}$ is the projection operator on the subspace of opened
Protonium formation channels, i.e. the Protonium states with principal
quantum number N $\leq $24. It is easy to see, that the first component
asymptotically describes elastic channel, while the second describes all the
inelastic channels. The equation system, obtained for the wave- function
components $\Phi _{1}$ and $\Phi _{2}$ is transformed into the one-channel
Schrodinger equation, governing the $H\bar{H}$ relative motion wave-function 
$\chi (R)$: 
\[
\left[ -{\frac{1}{M}}\partial _{R}^{2}+\widehat{W_{eff}}%
+V_{loc}(R)+U_{nucl}(R)-E+2\varepsilon _{B}\right] \chi (R)=0 
\]
where M is the \={p} mass, E is the $H\bar{H}$ center of mass energy, $%
\varepsilon _{B}$ is the H-atom Bohr energy, $V_{loc}$ is the local part of
interaction, obtained within adiabatic approximation, $\widehat{W_{eff}}$-
nonlocal, complex operator, which accounts for the coupling with the Pn+Ps
formation channels, U$_{nucl}$ is a nuclear potential of Woods-Saxon type,
fitted to describe $p\overline{p}$ nuclear scattering length. Effective
interaction $\widehat{W_{eff}}$ is calculated by solving the coupled
equations system for propagator (Green function) of Pn+Ps formation
channels. In practical calculations only the channels with Pn principal
quantum number $21\leq N\leq 24$ , Pn angular momentum $L\leq 1$, and Ps
quantum numbers $n\leq 2,l\leq 1$ were included. It was checked that
coupling with other channels is relatively small. In particular all channels
with production of unbound $e^{-}e^{+}$ were excluded due to their vanishing
contribution.

\section{RESULTS}

The $H\bar{H}$ scattering length is found to be $a^{H\overline{H}%
}=(6.1-i2.7) $ a.u. It was found to be an oscillating function of reduced
mass of $H\bar{H}$ thought of as a free parameter (see Fig.\ref{Fig1}.) The
immediate consequence of such a behavior of the scattering length is a
strong isotope effect. In particular, the $D\overline{H}$ scattering length
was found to be $a^{D\overline{H}}=(15.0-i11.6)$ a.u. The annihilation
cross-sections for $H\bar{H}$ and $D\bar{H}$ systems are presented on Fig.%
\ref{Fig2}. It should be mentioned that the semiclassical models show no
sign of oscillating behavior of observables. The extrapolated to low
energies values of semiclassical cross-sections depend on velocity $v$ like $%
(1/v)^{2/3}$ in contradiction with quantum mechanical $1/v$-law and
significantly smaller than the calculated quantum cross-section.

An oscillating behavior of scattering length as a function of reduced mass
is explained by existence of nearthreshold quasibound states \cite{C},
generated by van der Waals long range attraction. Appearance of the new
nearthreshold states with increasing of the reduced mass of the system
results in a strong enhancement of annihilation cross-section. The energies,
inelastic widths and mean radius of several such states are shown in Table %
\ref{Table1}.

Taking in mind, that main properties of observables are governed by $%
-C_{6}/R^{6}$ tail of local potential, while the short range interaction,
including effective potential, only modifies the spectrum of the
nearthreshold states, we can get the analytical expression for the $H\bar{H}$
scattering length: 
\[
a^{H\overline{H}}=\left( 2MC_{6}\right) ^{1/4}\frac{\Gamma (3/4)}{2\sqrt{2}%
\Gamma (5/4)}(1+\cot (\pi /8+\delta +\frac{\sqrt{2MC_{6}}}{2}r_{0}^{-2}))
\]%
where C$_{6}$ is van der Waals constant, r$_{0}$ is the range of short range
interaction (r$_{0}\approx 1$ a.u.), $\delta $ is a complex phase shift,
produced by short range complex part of the potential. It is clear from the
above expression, that if Im$\delta \gg 1$ (so called black sphere limit)
the scattering length becomes smooth function of reduced mass M and van der
Waals constant C$_{6}$ and insensitive to any details of short range
interaction: 
\[
a^{H\overline{H}}=\left( 2MC_{6}\right) ^{1/4}\frac{\Gamma (3/4)}{2\Gamma
(5/4)}\exp (-i\pi /4)
\]%
Our numerical calculations show that we are far from the black sphere limit,
and the scattering length is indeed an oscillating function of M and C$_{6}$%
. We have also find it to be very sensitive to short range interaction
details. In particular, taking into account the nuclear potential, which
describes the nuclear absorption, significantly changes the value of
scattering length from $a^{H\overline{H}}=5.2-i1.8$ a.u. to its final value
of $a^{H\overline{H}}=(6.1-i2.7)$ a.u.


\section{CONCLUSION}

A coupled channels model describing the $H\overline{H}$ system at energies
less than $10^{-5}$ a.u. has been developed. The results thus obtained for
annihilation cross-section substantially differ from the low energy
extrapolations of the black sphere model and other classical or
semiclassical approaches.

The reaction dynamics is found to be determined by the existence of several
nearthreshold states. Such states are produced by van der Waals potential
and have inelastic widths due to the coupling with Protonium formation
channels. We predict strong isotope effect in $H\bar{H}$ and $D\bar{H}$
scattering. Our results seem to be in a qualitative agreement with recently
published results \cite{Fr} by P. Froelich et al.

\begin{table}[tbp]
\[
\begin{array}{lllll}
\hline
E_I & E_{II} & \bar{x}_{II} &  &  \\ \hline
-8.1\;10^{-6} & -6.3\;10^{-6} +i\;1.8\;10^{-5} &  &  &  \\ \hline
-1.9\;10^{-4} & -4.3\;10^{-4} -i\; 2.2\;10^{-4} & 4.6 &  &  \\ \hline
-2.9\;10^{-3} & -5.2\;10^{-3} -i\;1.4\;10^{-3} & 3.2 &  &  \\ \hline
-1.1\;10^{-2} & -2.8\;10^{-2} -i\;8.2\;10^{-3} & 1.6 &  &  \\ \hline
\end{array}
\]%
\caption{Energies, Auger widths and mean radii (a.u) of L=0 $H\bar{H}$
states. We denote by index I the results in van der Waals potential alone
and by index II those obtained in full interaction.}
\label{Table1}
\end{table}
\newpage 
\begin{figure}[tbp]
\caption{$H\bar{H}$ annihilation cross-section (E=$10^{-8}$ a.u.) versus
reduced mass}
\label{Fig1}
\end{figure}
\newpage 
\begin{figure}[tbp]
\caption{$H\bar{H}$ and $D\bar{H}$ annihilation cross-sections}
\label{Fig2}
\end{figure}

\end{document}